\documentclass[preprint,aps,preprintnumbers]{revtex4}

\usepackage{epsfig}
\usepackage{amssymb}
\usepackage{amsfonts}
\usepackage{subfigure}
\usepackage{float}
\newcommand{\noi}{\noindent}
\newcommand{\beq}{\begin{equation}}
\newcommand{\eeq}{\end{equation}}
\newcommand{\bdm}{\begin{displaymath}}
\newcommand{\edm}{\end{displaymath}}
\newcommand{\bea}{\begin{eqnarray}}
\newcommand{\eea}{\end{eqnarray}}

\begin{document}
\preprint{HU-EP-03/84}
\preprint{IFUP-TH-2003/45} 
\preprint{TRINLAT-03/08} 

\title{SO(3) vs. SU(2) Yang-Mills theory on the lattice: \\
an investigation at non-zero temperature
\footnote{Based on contributions by 
G. Burgio and M. M\"uller-Preussker at CONFINEMENT 2003, RIKEN, Tokyo
and LATTICE 2003, Tsukuba, Japan}}  

\author{A. Barresi} 
\affiliation{Dipartimento di Fisica, Universit\'a di Pisa e 
I.N.F.N. Sezione di Pisa, 
Via Buonarroti 2, 56127 Pisa, Italy \\
E-mail: barresi@df.unipi.it}
\author{G. Burgio}
\affiliation{School of Mathematics, Trinity College,
Dublin 2, Ireland \\
E-mail: burgio@maths.tcd.ie}
\author{M. M\"uller-Preussker}
\affiliation{Humboldt-Universit\"at zu Berlin, Institut f\"ur Physik, 
Newtonstr. 15, 12489 Berlin, Germany \\
E-mail: mmp@physik.hu-berlin.de}

\date{\today}

\begin{abstract}
The adjoint $SU(2)$ lattice gauge theory in 3+1 
dimensions with the Wilson plaquette action modified by a 
$\mathbb{Z}_2$ monopole suppression term is reinvestigated
with special emphasis on the existence of a finite-temperature
phase transition decoupling from the well-known bulk
transitions. 
\end{abstract}	

\maketitle	

\section{Introduction and motivation}

The evidence and our detailed understanding of the deconfinement phase 
transition in $SU(N)$ gauge theories at finite temperature 
mainly comes from lattice gauge theories (LGT) formulated 
in the fundamental representation \cite{McLerran,Kuti}.
For pure LGT the transition is associated with the spontaneous 
breaking of the global center $\mathbb{Z}_N$ symmetry 
\cite{Polyakov,Susskind}: 
  $$ U_4(\vec{x},x_4) \longrightarrow  z \cdot U_4(\vec{x},x_4), 
      \quad z \in \mathbb{Z}_N
  \quad \mbox{for all} ~~\vec{x}~~ \mbox{at} ~~x_4 = \mbox{fixed}.$$
which leaves the lattice gauge action invariant but flips
the Polyakov loop variables 
\beq
L_F(\vec{x})=\frac{1}{N_c} \mathrm{Tr}_F 
                           \prod_{x_4=1}^{N_{\tau}} U_4(\vec{x},x_4) 
\eeq
as $L_F \leftrightarrow z L_F.$  As a consequence the standard order 
parameter for the deconfinement transition is defined as
\beq 
< ~|L_F|~ > =  \left< ~\left| \frac{1}{N_{\sigma}^3}
            \sum_{\vec{x}} L_F(\vec{x}) \right|~ \right>\,, 
\eeq
where the ensemble average is taken with the Boltzmann distribution
represented by the lattice-discretized path integral with periodic
boundary conditions for the gauge fields in the imaginary time direction
$x_4$. The above mentioned global symmetry breaking mechanism 
provides a close analogy to spin models. In particular, the universality 
class of $SU(2)$ LGT is that of the 3d Ising model \cite{Svetitsky}.                              
On the other hand the origin of quark and gluon confinement as well as 
of the occurence  of the finite-temperature phase transition 
has been seen in the condensation of topological excitations 
like Abelian monopoles \cite{monopoles} and center vortices \cite{vortices}. 

Lattice gauge theories can be formulated in different group
representations of the gauge fields, e.g. in the {\it center blind} 
adjoint representation. In this case (extended) vortices and Abelian
monopoles are still present, but the mechanism of spontaneous 
$\mathbb{Z}(N)$ breaking is obviously not realized. Moreover, the
adjoint representation LGT's at strong coupling are strongly affected 
by bulk phase transitions \cite{Bhanot,Greensite} driven by lattice 
artifacts \cite{Halliday1}. A finite temperature transition - 
if it exists - seems to be completely overshadowed by these bulk 
transitions. 

Therefore, the question of universality in particular of the existence of
the finite temperature phase transition remains an important issue.
If the existence of this transition turns out to be independent of 
the group representation, then the question remains whether 
the driving mechanism related to the condensation of topological
excitations is the same.  

In the past this principally important question has been studied 
by several groups mainly in the case of the mixed
$~SU(2) - SO(3)=SU(2)/\mathbb{Z}_N$ theory realized with
the Villain action$^{11-17}$. 
Still we have not yet reached a completely satisfying answer.
Nevertheless, over the last years there has been an interesting 
progress$^{18-21}$
worth to be reviewed at this conference.

In the following Section 2 we shall shortly review $SU(2)$ lattice gauge 
theories with different mixed fundamental-adjoint actions. In Section 3 we 
introduce the center-blind model we have further investigated, i.e. 
the adjoint representation Wilson lattice action with a $\mathbb{Z}_2$ 
monopole suppression term.
In Section 4 we discuss the results of our investigations based on 
twist variables, the fundamental Polyakov loop distributions as well as
on the Pisa disorder operator providing evidence for the existence
of a distinct finite-temperature transition in the center-blind theory.
Our conclusions are drawn and an outlook is given in Section 5.

\section{SU(2) lattice gauge theories with mixed fundamental-adjoint action}

Among the ``first day'' lattice gauge theory models were also those 
with a mixture of different group representations
for the plaquette contribution, e.g. for $SU(2)$ \\
$-$ the Wilson-type mixed action \cite{Bhanot}
\bea
S &=& \beta_{A}\sum_{P}\Bigg(1-\frac{1}{3}\mathrm{Tr}_{A}U_{P}\Bigg)
   +\beta_{F}\sum_{P}\Bigg(1-\frac{1}{2}\mathrm{Tr}_{F}U_{P}\Bigg) \,, \\
\frac{1}{g^2} &=& \frac{\beta_F}{4}+2\frac{\beta_A}{3} \nonumber
\eea
$-$ the Villain-type mixed action \cite{Halliday1}
\beq
S = \beta_{V}\sum_{P}\Bigg(1-\frac{1}{2}\sigma_{P}\mathrm{Tr}_{F}U_{P}\Bigg)
   +\beta_{F}\sum_{P}\Bigg(1-\frac{1}{2}\mathrm{Tr}_{F}U_{P}\Bigg)\,,
\eeq
where $\sigma_{P}=\pm 1$ is an auxiliary dynamical 
$\mathbb{Z}_2$ plaquette variable.

\begin{figure}[ht]  
\vspace{0.5cm}
\centerline{\epsfxsize=3.9in \epsfysize=1.5in \epsfbox{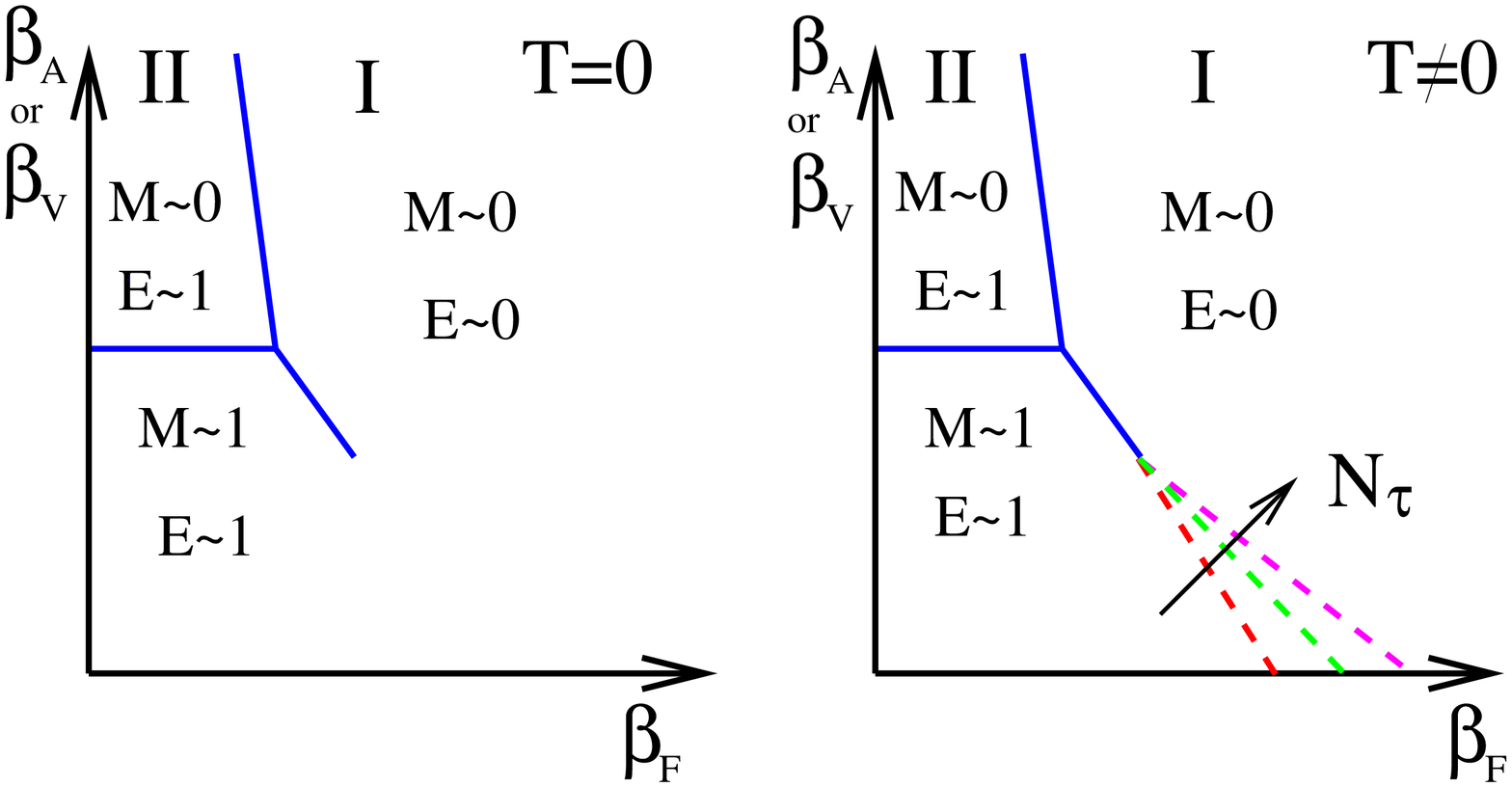}}   
\caption{
Schematic phase structure at $T=0$ and $T>0$. 
\label{fig.1}}
\end{figure}

The non-trivial phase structure with first order bulk transitions 
(see Fig. \ref{fig.1})
is governed by lattice artifacts: $\mathbb{Z}_2$ magnetic monopoles 
and  electric vortices the densities of which can be defined as follows
($N_c$ and $N_l$ being the number of 3-cubes and lattice links, 
respectively)\cite{Halliday2} 
\bea   
M&=&1-\langle\frac{1}{N_{c}}\sum_{c}\rho_{c}\rangle\,, \qquad 
     \rho_{c}=\prod_{P\in\partial c}\sigma_{P} ~~~\mbox{or}~~~ 
              \prod_{P\in\partial c}\mathrm{sign}(\mathrm{Tr}_F U_{P}) \\
E&=&1-\langle\frac{1}{N_{l}}\sum_{l}\rho_{\,l}\rangle\,, \qquad
     \rho_{\,l}=\prod_{P\in\hat{\partial}\,l}\sigma_{P}
      ~~~\mbox{or}~~~
     \prod_{P\in\hat{\partial}\,l}\mathrm{sign}(\mathrm{Tr}_F U_{P})\,.
\eea
These lattice excitations can be suppressed by modifying the action with
suppression terms like \cite{Halliday2,Datta2} 
\beq
\lambda_{V}\sum_{c}(1-\rho_{c})\,, \hspace{1cm} 
\gamma_{V} \sum_{l}(1-\rho_{\,l})\,.
\eeq
For the Villain-type action the equivalence between $SO(3)$ and $SU(2)$ 
has been proven in the limit of complete $\mathbb{Z}(2)$ monopole
supression $~\lambda_V \to \infty~$ for $(\beta_F=\gamma_V=0)~$
$^{22-25,18}$
in the following form
\bdm
\sum_{\mathrm{twist~sectors}}~Z_{SU(2)}~
\equiv A~\sum_{\sigma_P=\pm 1} \int DU e^{\beta_{V}\sum_{P}\sigma_{P}
   \frac{1}{2}\mathrm{Tr}_{F}U_{P}} 
   \prod_c \delta(\prod_{P\in\partial c}\sigma_{P} - 1)
\edm
where on the l.h.s. the twist sectors are imposed by 
twisted boundary conditions
$U_{\nu}(x+L_{\mu}) = z_{\mu\nu} U_{\nu}(x), 
\quad z_{\mu\nu} \in \pi_1 [SU(2)/\mathbb{Z}_2]=\mathbb{Z}(2).$
On the r.h.s. the twist sectors are dynamically encountered, 
under circumstances separated by large barriers.  
  
The case $~T \ne 0$ has been mostly studied with the modified 
Villain action but always with a non-vanishing admixture of the 
fundamental representation ($\beta_F \ne 0$).
Lines of a finite-$T$ phase transition presumely of second order 
have been found in the $\beta_V - \beta_F$ plane for 
$\lambda_V \ge 1$ and $\gamma_V \ge 5$
\cite{Datta2}. Above the finite-$T$ transition the adjoint Polyakov line 
$\langle L_{A}\rangle$  has been seen trapped into metastable states
\cite{Cheluvaraja,Datta1}
\beq
\langle L_A \rangle \longrightarrow 
\left\{ \begin{array}{c} 1 \\ -\frac{1}{3} \end{array} \right. \qquad \mbox{as} 
\quad \beta_V \to \infty
\eeq
\noi
Jahn and de Forcrand \cite{Forcrand1} related the negative $L_A$ states 
to non-trivial twists. For demonstrating this they introduced $SO(3)$ -- 
i.e. center-blind -- twist variables
\beq
z_{\mu\nu} \equiv \frac{1}{N_{\rho} N_{\sigma}} \sum_{\rho \sigma} 
    \prod_{P \in \mu,\nu -\mathrm{plane}} \mathrm{sign}~
    (\mathrm{Tr}_F \,U_P) \in [-1,+1] \,, \quad \epsilon_{\mu\nu\rho\sigma} = 1.
\label{eq.twists}
\eeq       
\noi
The $z_{\mu\nu}$'s measure the $\mathbb{Z}(2)$ fluxes through $\mu\nu$-planes.
Then the state 
\beq    
L_F=0 ~\Longleftrightarrow~L_A 
\equiv \frac{1}{3} \mathrm{Tr}_A L= -\frac{1}{3}\,, 
\hspace{1.0cm} \mathrm{Tr}_{A} L = (\mathrm{Tr}_{F} L)^2-1 
\eeq
is related to electric twist $ z_{i,4} = -1, \qquad i=1,2,3\,.$                 

Having these observations in mind we are going now to check
and to illustrate this scenario for a center-blind modified adjoint 
Wilson action. We ask how to establish a finite $T$ transition 
for the center-blind theory and what r\^ole do play the different 
twist sectors in this case.

\section{Adjoint SU(2) model with $\mathbb{Z}_2$ monopole suppression}

\begin{figure}[ht]  
\vspace{0.5cm}
\centerline{\epsfxsize=3.9in \epsfysize=1.5in \epsfbox{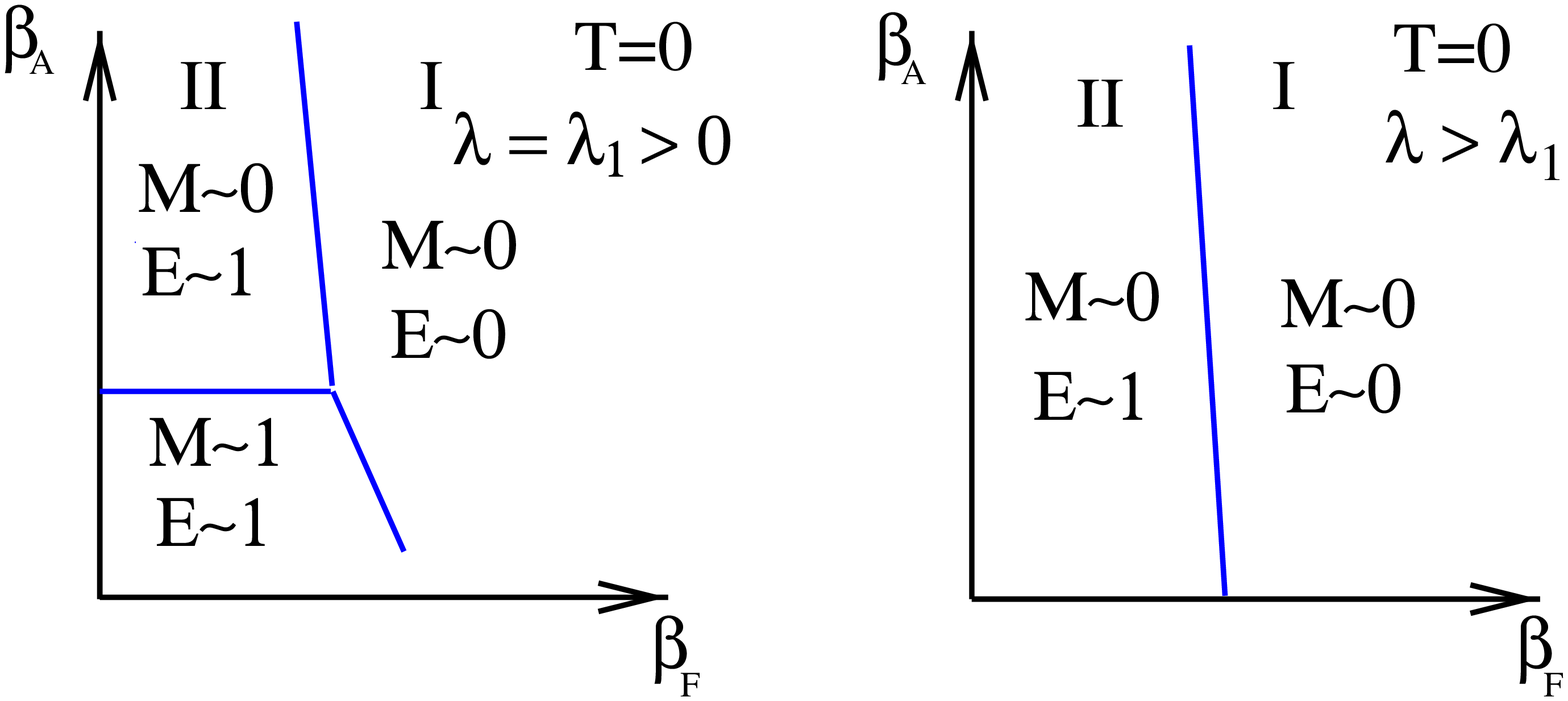}}   
\caption{
Schematic $\beta_F-\beta_A$ phase diagram for varying $\lambda$ at $T=0$. 
\label{fig.2}}
\end{figure}

In our investigations we have considered the Wilson plaquette action 
with link variables $U_{\mu}(x) \in SU(2)$
\beq
S  =  \beta_{F} \sum_{P}\Bigg(1-\frac{1}{2}\mathrm{Tr}_{F}\, U_{P}\Bigg)
     +\beta_{A} \sum_{P}\frac{4}{3}\Bigg(1-\frac{1}{4}(\mathrm{Tr}_{F}\, 
     U_{P})^2 \Bigg) +\lambda \sum_c (1-\rho_c) 
\eeq     
where
$\rho_{c} = \prod_{P \in \partial c} \mathrm{sign} \, \mathrm{Tr}_F \, U_{P}.$ 
For $\beta_F = 0$ the action $~S~$ becomes center-blind  
$$U_{\mu}(x)\rightarrow -1 \cdot U_{\mu}(x) ~~\Longrightarrow~~
               \rho_{c} \rightarrow \rho_{c}\,.$$               
Fig. \ref{fig.2} shows the phase diagram in the $\beta_F-\beta_A$-plane
for varying chemical potential $\lambda$ for $T=0$.
Obviously, the suppression of $\mathbb{Z}_2$ monopoles ($\lambda>0$) 
shifts the horizontal line down to smaller $\beta_A$-values. At a first 
glance the phase II seems to be disconnected from phase I 
(the ordinary confinement phase) in the range
$0 \le \lambda \le 1$. But see the discussion further below.

\begin{figure}[ht]
\begin{center}
\includegraphics[angle=0,width=0.6\textwidth]{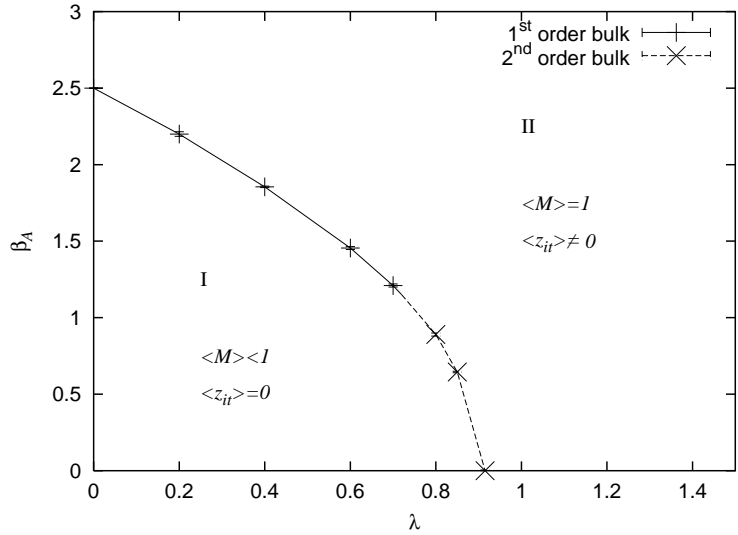}
\end{center}
\caption{
The bulk transition line in the $\beta_A-\lambda$ plane as seen for 
the lattice size $12^3 \times 4$.
\label{fig.3}}
\end{figure}

If we put $\beta_F=0$ the emerging $\beta_A$-$\lambda$ diagram 
looks as shown in Fig. \ref{fig.3}.
\noi
Phase I -- which at $\beta_F \ne 0$ is connected with the ordinary 
confinement phase -- is characterized by a non-zero $\mathbb{Z}(2)$ monopole
density and by twist variables (\ref{eq.twists}) fluctuating close
to zero. On the contrary in phase II the monopoles become suppressed
and the twist variables (meta)stable at $\pm 1$. 
The phase transition line has been established
by studying the average plaquette, the adjoint Polyakov loop variable
$<L_A>$, the average density $M$ of $\mathbb{Z}(2)$ monopoles and 
additionally the twist variables (\ref{eq.twists}) as well as their 
'susceptibility'
\beq
\chi_{\mathrm{twist}}= N_{\sigma}^3 \cdot (\langle\tilde{z}^2\rangle-
\langle{z}\rangle^2)
\label{eq.susc}
\eeq
with $~\tilde{z}\equiv\frac{1}{3}(|z_{xt}|+|z_{yt}|+|z_{zt}|).$ 

We found out that the bulk transition line 
in a certain range $0 < \lambda < \lambda_c$  (with $\lambda_c \simeq 0.7$
for the lattice size $12^3 \times 4$) shows 
a discontinuous behaviour of the monopole density $M$, of the 
average plaquette as well as of the adjoint Polyakov loop  
across the transition. At the same time tunneling between different 
twist sectors becomes strongly suppressed when passing the border 
from  phase I to phase II. Along this line
as long as  $\lambda < \lambda_c$ the twist sectors are clearly related 
to metastable states of the adjoint Polyakov loop. We interpret the 
transition likely to be a first order transition in this range.

\begin{figure}[ht]
\begin{center}
\includegraphics[angle=0,width=0.5\textwidth]{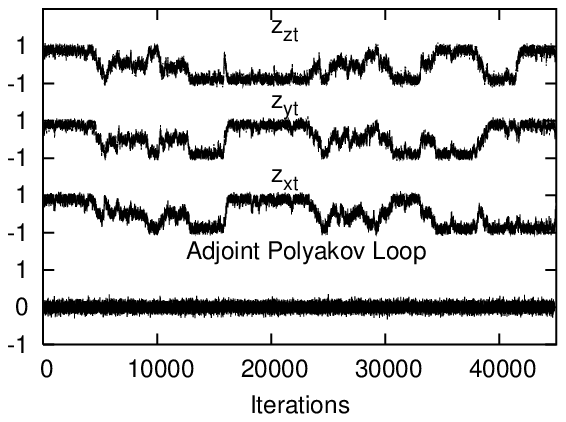}
\includegraphics[angle=0,width=0.48\textwidth]{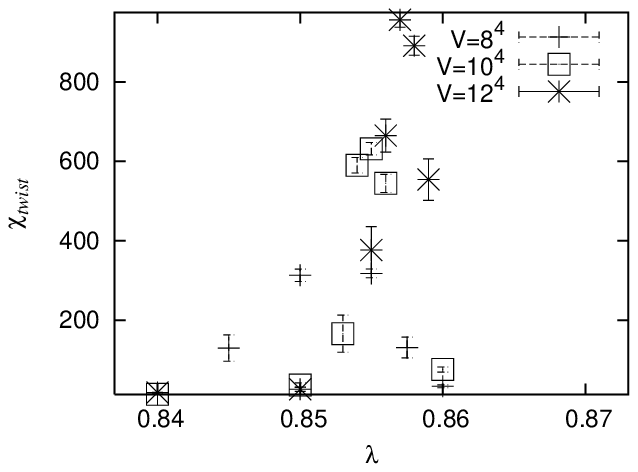} 
\caption{Electric twist histories at $\beta_A=0.65, 
\lambda=0.858$ for $12^4$ (left) and twist susceptibility for 
$\beta_A=0.65$ at varying $\lambda$ and lattice sizes 
$8^4, 10^4$ and $12^4$ (right). 
\label{fig.4}}
\end{center}
\end{figure}

On the contrary, for $\lambda_c < \lambda < 0.95$ the monopole density 
$M$ and the average plaquette turn out to behave smoothly across the bulk
transition line. For the average adjoint Polyakov loop one finds 
$<L_A> \simeq 0$ on both sides of the line, i.e. phase II in this range
seems to be also confining. This already is an indication that for larger 
$\lambda$-values and increasing $\beta_A$ there should be a further
(more or less horizontal) transition line hopefully behaving as a finite 
temperature transition. The 'tricritical' value $\lambda_c$ seems to 
indicate the position, where this additional line might join the bulk 
transition. We come back to this question in the next section.
The bulk transition itself in the range $\lambda_c < \lambda < 0.95$ is 
visible because of the enhanced tunneling between different twist sectors. 
This we have observed for all lattice sizes (so far up to $V=12^4$). The 
twist susceptibility (\ref{eq.susc}) has a strong peak which increases with 
increasing lattice size (see Fig. \ref{fig.4}). 
A rough finite-size scaling test shows the transition to resemble 
the 4D Ising one and, therefore, seems to be of second order.

\section{Evidence for a finite-temperature transition}

We observed that for sufficiently large chemical potential 
$~\lambda \ge 1.0~$ tunneling between twist sectors becomes 
completely suppressed. We decided to run the simulations 
in this range within fixed twist sectors (mostly the trivial one). 
That is, we suppress the generation or annihilation of extended 
vortices (up to numbers modulo two).

\begin{figure}[ht]
\begin{center}
\includegraphics[angle=0,width=0.7\textwidth]{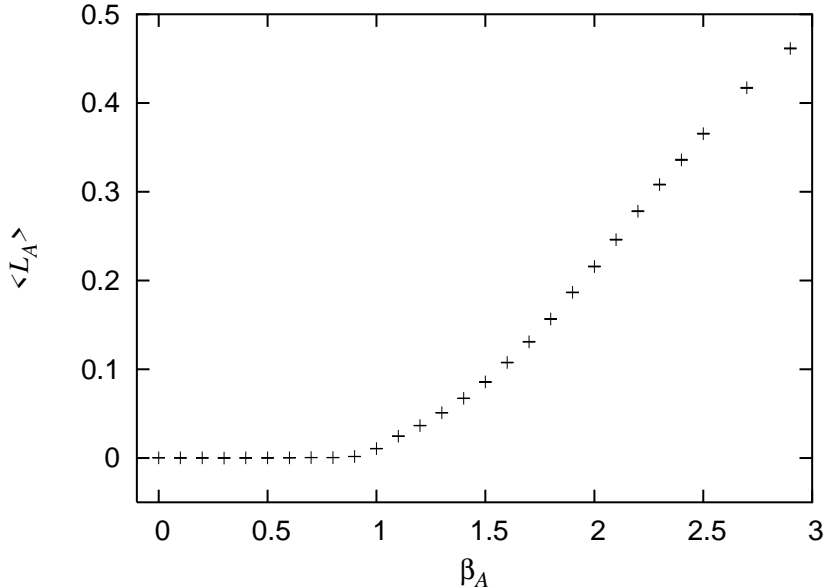}
\end{center}
\caption{Adjoint Polyakov loop average as a function of 
$\beta_A$ at $\lambda=1.0$ for lattice size $V=4\times 12^3$.
\label{fig.5}}
\end{figure}

\begin{figure}[ht]
\begin{tabular}{ccc}
$\beta_A=0.9$ & $\beta_A=1.2$ & $\beta_A=1.4$ \\
\includegraphics[angle=0,width=3.0cm]{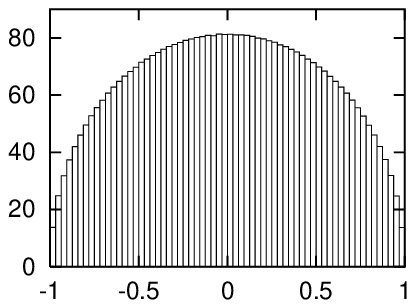} &
\includegraphics[angle=0,width=3.0cm]{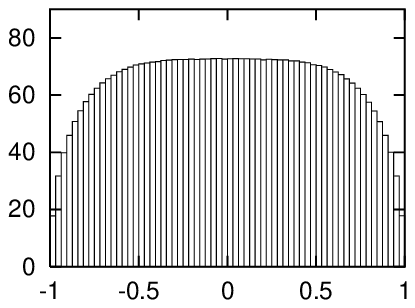} &
\includegraphics[angle=0,width=3.0cm]{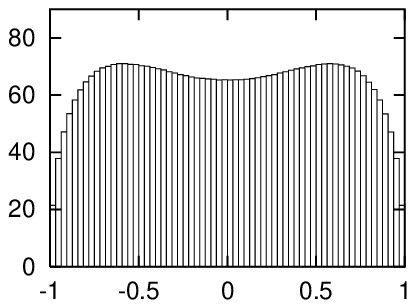} \\~\\
\end{tabular} 
\begin{tabular}{ccc}
$\beta_A=1.4$ & $\beta_A=1.6$ & $\beta_A=1.8$ \\
\includegraphics[angle=0,width=3.0cm]{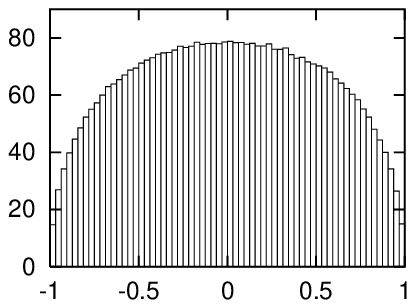} &
\includegraphics[angle=0,width=3.0cm]{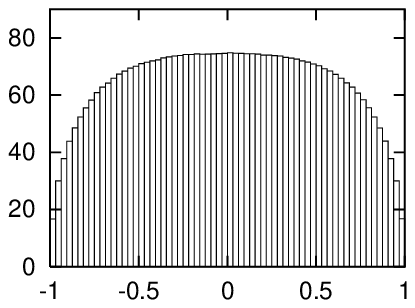} &
\includegraphics[angle=0,width=3.0cm]{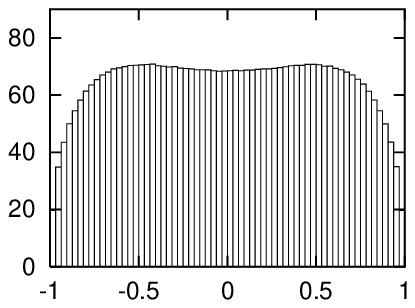}
\end{tabular}
\caption{Frequency distributions of the local fundamental Polyakov 
loop variable $L_F(\vec{x}).$ Upper row for lattice size $V = 4 \times 16^3$,
lower row for $V= 6 \times 16^3$, all for  $\lambda=1.0$ and zero twist.
\label{fig.6}}
\end{figure}

For fixed $\lambda$ we carried out measurements with varying
$\beta_A$. The adjoint Polyakov loop average $\langle L_A \rangle$ 
as a function of $\beta_A$ in the zero twist sector is drawn in 
Fig. \ref{fig.5}. Its rise clearly indicates a transition close
to $\beta_A \simeq 1.$
\noi
Frequency distributions of the local fundamental Polyakov loop variable 
$L_F(\vec{x})$ are plotted for two temporal lattice extensions in 
Fig. \ref{fig.6}. One sees that the shape of the distributions clearly 
changes from a phase, where they peak at zero, to a phase, where they
have two symmetric maxima away from zero. The critical $\beta_A$, where
two maxima just occur, changes to larger values as $N_t$ is increasing from
$4$ to $6$ in agreement with scaling. Therefore, we can conclude that
a finite-temperature transition is really seen. 
In order to check the existence of the transition at finite $T$ 
with an independent measurement we have computed also the  
Pisa disorder parameter\cite{Giacomo1,Giacomo2}
which on the basis of the dual superconductor model \cite{monopoles}
allows to test for a condensate of Abelian monopoles in the 
confinement phase. The order operator for the condensation of magnetic 
charges $\mu(t)$ is defined by modifying the action at a given time-slice 
$t$ by a classical Dirac monopole field insertion $\Phi$ 
\cite{Giacomo1,Giacomo2}
\bea
\mu(t) & = & \exp(-\beta\Delta S(t)) \\
\Delta S(t) & = & \frac{1}{2}
     \sum_{i,\vec{x}} \,\mathrm{Tr} \,[U_{i4}(\vec{x},t)-U_{i4}'(\vec{x},t)] \\
U_{i4}'(\vec{x},t) & = & 
U_{i}(\vec{x},t)\, \Phi_i(\vec{x} + \hat{i},\vec{y}) \, U_{4}(\vec{x}+\hat{i},t)\,
 U^{\dagger}_{i}(\vec{x},t+1)\,U^{\dagger}_{4}(\vec{x},t)
\eea
The operator can be generalized to the adjoint ($SO(3)$) action case.
In the thermodynamic limit one expects
\bdm
\langle\mu\rangle~~~~\left\{\begin{array}{ll}
      \neq 0 & ~~~T<T_c \\
      =0     & ~~~T>T_c
      \end{array}\right.
\edm
In practice, one measures instead
\bea
\rho=\frac{d}{d\beta}\log\langle\mu\rangle=
-\langle S+\Delta S\rangle|_{S+\Delta S}+\langle S\rangle|_{S}.
\eea
The experience for the $SU(2)$ and $SU(3)$ cases tells that the latter
quantity exhibits a sharp dip signalling the deconfinement phase transition 
to be driven by the breaking of a dual magnetic symmetry.
    
\begin{figure}[ht]
\begin{center}
\begin{tabular}{ccl}
    $\lambda=0.7$ & $\hspace{-0.7cm}\lambda=0.85$ & $\hspace{.2cm}\lambda=1.0$ \\ \hspace{-0.7cm}
\includegraphics[angle=0,width=.38\textwidth]{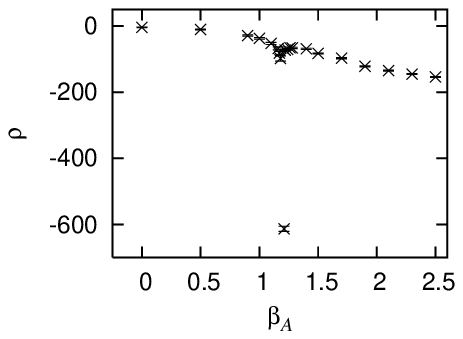}&
\includegraphics[angle=0,width=.38\textwidth]{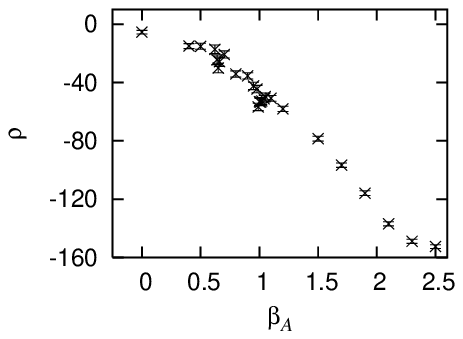}&
\includegraphics[angle=0,width=.38\textwidth]{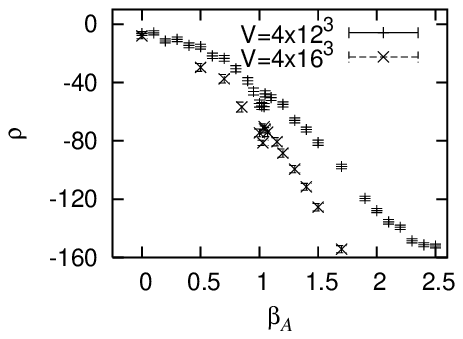}
\end{tabular}
\end{center}
\caption{
Pisa disorder parameter as a function of $\beta_A$ 
for various $\lambda$ values (lattice size $V = 4 \times 12^3$).
The right most figure shows additionally the case with a larger 
volume $V=4 \times 16^3$.	
\label{fig.7}}	
\end{figure}

Some of our results for the $SO(3)$ case are collected in Fig. \ref{fig.7}.

\noi
The computations show that for $\lambda = 0.7$ the bulk phase transition 
I $\leftrightarrow$ II is correctly localized. For increasing $\lambda$, e.g.
at $\lambda=0.85$, we see simultaneous, distinct signals for the bulk 
and the finite-$T$ transitions. At even larger chemical potential ($\lambda=1.$) 
we find only a signal for the finite-$T$ transition.
The localizations of the bulk and finite temperature transitions agree 
reasonably with those obtained with the other methods mentioned before.
The universality class of the transition to be determined e.g. via 
the critical indices has still to be investigated. 

Having convinced ourselves that there is a finite temperature transition
also in the completely center-blind adjoint $SU(2)$ LGT we are tempted to ask
whether there is a spontaneous breaking of an appropriate global symmetry.
If yes, what is the corresponding order parameter?

For $SO(3)$ the question means, whether there is a globally defined operator 
flipping the adjoint Polyakov loop $~L_A = 1~\Leftrightarrow~L_A=-\frac{1}{3}$ while
leaving the action invariant. There is an approximate solution based on the dual 
superconductor picture assuming that the long-distance properties are carried by 
Abelian degrees of freedom.

Let us consider flip operators with $P \in SU(2)$ or $SO(3)$ satisfying the 
conditions
$~~P^2 = \pm \mathbb{I},~~P^{\dagger}=\pm P~~$. The only ones fulfilling these
conditions are for
\bea 
SU(2):~~~P &=& \pm \mathbb{I}_2,~~\hat{P}=~\pm i \vec{n} \cdot \vec{\sigma}\,, 
\nonumber\\
SO(3):~~~P &=& + \mathbb{I}_3,~~\hat{P}=~\mathbb{I}_3 + 2 (\vec{n} \cdot \vec{T})^2, 
\qquad (~n^2 = 1~).
\nonumber
\eea

\begin{figure}[ht]
\begin{center}
\includegraphics[angle=0,width=0.48\textwidth]{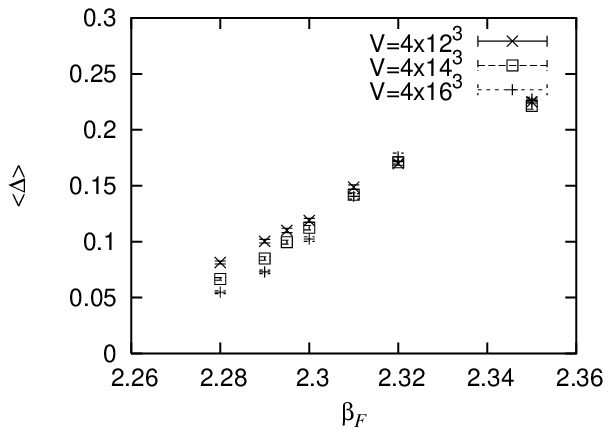}
\includegraphics[angle=0,width=0.48\textwidth]{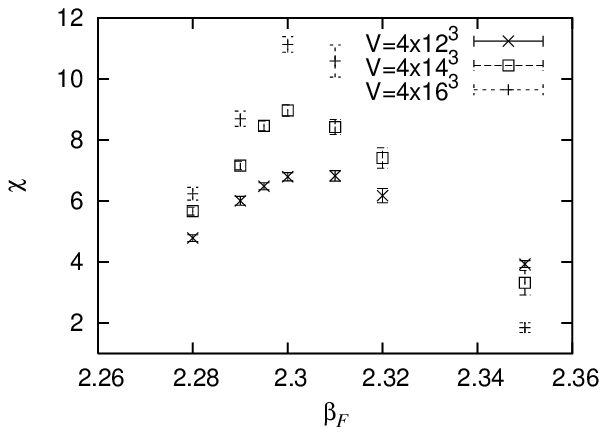}
\end{center}
\caption{
`Order parameter' $\langle \Delta \rangle$ (left) and its susceptibility $\chi$ 
(right) versus $~\beta_F$ for $\beta_A=0.0,~\lambda=0.0$.	
\label{fig.8}}
\end{figure}

\noi
Let us assume the $SU(2)$ or $SO(3)$ fields to be Abelian (diagonal) with respect 
to a fixed `isospin' $~\vec{n}$-direction, then $~\hat{P}~$ applied to a given time 
sheet, indeed, leaves the action invariant, and the Polyakov loop is 
flipped to the other state  (if its value is different from zero).
In practice, this is realized only approximately by fixing the 3D maximally 
Abelian gauge (MAG) on each time slice $x_4$ separately. This is
achieved by maximizing the gauge functional with respect
to gauge transformations $~g$, e.g. for $SU(2)$:
\beq
F_{x_4}(g) = \sum_{\vec{x}} \sum_{i=1,2,3} 
             \mathrm{Tr}\left((\vec{n} \cdot \vec{\sigma})~U_i^g (\vec{x},x_4)~ 
             (\vec{n} \cdot \vec{\sigma})~(U_i^g(\vec{x},x_4))^{\dagger}\right). 
\eeq
The `order parameter' is then defined as 
\beq
\langle \Delta_{F,A} \rangle = 
   c_{F,A} ~\langle~| L_{F,A} - \tilde{L}_{F,A} |~\rangle\,,
\eeq   
where 
\beq
\qquad \tilde{L}_{F,A} = \frac{1}{N_{\sigma}^3}\sum_{\vec{x}}
\mathrm{Tr}_{F,A} \left( \hat{P} \,\prod_{x_4=1}^{N_{\tau}} U_4(\vec{x},x_4) \right).
\eeq 
The `order parameter' should work as for the standard $SU(2)$ case as well as
for $SO(3)$. We have checked this. Indeed, for $SU(2)$ we obtained the results
drawn in Fig. \ref{fig.8} for the order parameter itself as well as for its
susceptibility.
\noi
The susceptibility develops a peak around $\beta_F=\beta_c \simeq 2.3$ 
as it should be. The corresponding Binder cumulant computed for varying 
3-volume has been seen to provide intersecting lines at the same 
$~\beta_F$ value.
If one computes the corresponding `order parameter' $\Delta$ for the adjoint $SO(3)$ 
action case at $\lambda=1.$ and in the zero-twist sector one gets a similar 
picture as the left one in Fig. \ref{fig.8}. 

\section{Conclusions and outlook} 

\begin{figure}[ht] 
\begin{center}
\includegraphics[angle=0]{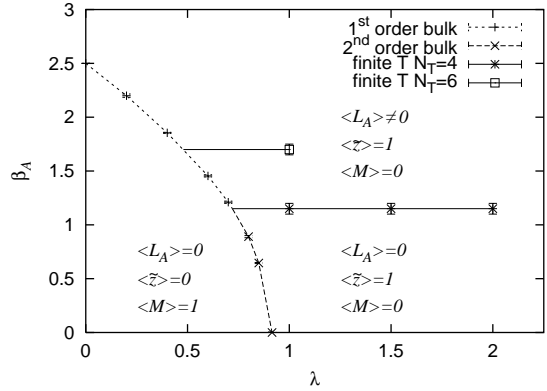}
\end{center}
\caption{
Phase diagram of the center-blind $SO(3)$ model with $\mathbb{Z}(2)$ 
monopole suppression term. The horizontal lines indicate the position
of the finite temperature transition for varying time extension 
$N_t=4, 6$.	
\label{fig.9}}	
\end{figure}

We summarize our studies by drawing the phase diagram for the modified adjoint 
$SU(2)$ theory at $T > 0$ as shown in Fig. \ref{fig.9}.
We see the existence of a finite temperature transition decoupling from
the bulk transition at a position which scales in $\beta_A$ as one would 
expect. The finite temperature transition was also localized with the
help of the Pisa disorder parameter indicating that the dual superconductor 
scenario seems to work also for $SO(3)$. Whether also the determination
of the free energy of an extended center vortex (see references 
\cite{Kovacs2,Forcrand2}) would also point to the same result remains to be seen.   
Our numerical study presented here is in many respects still preliminary. Larger
lattices, the use of an ergodic algorithm allowing to enhance tunneling
between different twist sectors and the determination of critical indices
in order to determine the universality class require more extensive 
computations, until we can draw final conclusions.  


\end{document}